\pdfoutput=1
\documentclass{PoS}
\usepackage{bm}        
\usepackage{amssymb}   
\usepackage{tikz}
\usepackage{amsmath}    

\title{Non-Local effective SU(2) Polyakov-loop models from inverse Monte-Carlo methods}

\ShortTitle{Non-Local SU(2) Polyakov models from IMC}
        
\author{\speaker{Bardiya Bahrampour}\thanks{This work was supported by the Helmholtz International Center for FAIR within the LOEWE initiative of the State of Hesse.}\\
        Institute for Theoretical Physics, Justus-Liebig University Giessen, Germany\\
        E-mail: \email{bardiya.bahrampour@physik.uni-giessen.de}}        
        
\author{Bj\"{o}rn Wellegehausen\\
	Institute for Theoretical Physics, Justus-Liebig University Giessen, Germany\\
        E-mail: \email{bjoern.Wellegehausen@theo.physik.uni-giessen.de}}        

\author{Lorenz von Smekal\\
        Institute for Theoretical Physics, Justus-Liebig University Giessen, Germany\\
        E-mail: \email{lorenz.smekal@theo.physik.uni-giessen.de}}

\abstract{The strong-coupling expansion of the lattice gauge action
  leads to Polyakov-loop models that effectively describe gluodynamics
  at low temperatures, and together with the hopping expansion of the
  fermion determinant provides insight into the QCD phase diagram at
  finite density and low temperatures, although for rather heavy
  quarks. At higher temperatures the strong-coupling expansion breaks
  down and it is expected that the interactions between Polyakov loops
  become non-local. Here, we therefore test how well pure SU(2)
  gluodynamics can be mapped onto different non-local Polyakov models
  with inverse Monte-Carlo methods. We take into account Polyakov
  loops in higher representations and gradually add interaction terms
  at larger distances. We are particularly interested in
  extrapolating the range of non-local terms in sufficiently large
  volumes and higher representations. We study the characteristic
  fall-off in strength of the non-local couplings with the interaction
  distance, and its dependence on the gauge coupling in order to
  compare our results to existing proposals for non-local effective
  actions.}

\FullConference{34th annual International Symposium on Lattice Field Theory\\
		24-30 July 2016\\
		University of Southampton, UK}

\begin{document}

\section{Introduction}
Due to the sign problem, QCD at finite density is still a challenge, a
number of different methods is currently actively being explored. 
One way to get further insight into the QCD phase diagram is to use
effective theories in which the sign problem is either absent or weak
enough to be dealt with \cite{1,2,3,4,5}. The deconfinement 
phase transition of a pure Yang-Mills theory in $d$ dimensions is determined  
by the dynamics of Polyakov loops. 
From the arguments by Svetitsky and Yaffe \cite{Yaffe2} it shares its
universal behavior with a spin model in $d-1$ dimensions, and 
Polyakov-loop models are excellent candidates of effective field
theories to describe this behavior. They can be derived in
strong-coupling expansions by integrating out the spatial links.
Unfortunately, however, with increasing temperature the strong-coupling
expansion eventually breaks down. In particular, local Polyakov-loop
models typically fail to describe the full SU(3) Yang-Mills theory
when the temperature approaches the phase transition.
Therefore, one is left with non-local Polyakov-loop models whose
effective couplings need to be mapped to the full theory in other ways. 

Here we restrict to the simpler case of SU(2) gauge
theory in which the fermion determinant at finite density remains real 
because of the pseudo-reality of its gauge group. In a first instance
we investigate different types of non-local Polyakov-loop models
for the pure gauge theory as motivated from the strong-coupling
expansion with and without resummations of higher-order terms. 
Because an analytical derivation of all the resulting terms is in
general not feasible, we will use the inverse Monte-Carlo 
method together with geometric Ward identities to fix the couplings of
our effective theories \cite{wipf}. Taking into account only Polyakov
loops that wind around the temporal direction once, we can write the
effective Polyakov-loop action as 
\begin{align}
S_{\text{lin}}=\sum_{p}\sum_{r^2\geq 1}\sum_{<i,j>=r^2}\lambda_{p,r^2}
\chi_{p,i}\chi_{p,j}\, ,
\label{linansatz}
\end{align} with lattice indices of sites $i$ and $j$ at distances
$r$, coupling constants $\lambda_{p,r^2}$ and Polyakov loops
$\chi_{p,i}$ in represenations $p$ given by their Dynkin labels. This
leads to the first class of effective theories which we refer to as
the {\em linear} Polyakov-loop models. Alternatively one may rearrange
classes of higher-order terms arising in the strong-coupling expansion and
perform partial resummations 
to motivate an ansatz for an effective theory with
logarithmic terms in the action \cite{resummation},
\begin{align}
S_{\text{log}}=-\sum_{p}\sum_{r^2\geq 1}\sum_{<i,j>=r^2}\ln\left(1+g_{p,r^2}
\chi_{p,i}\chi_{p,j}\right) \, ,
\label{logansatz}
\end{align}
with coupling contants $g_{p,r^2}$, which we refer to as the {\em
  logarithmic} Polyakov-loop models.

\section{Inverse Monte-Carlo Method, Geometric Ward-Identities and DSE's}
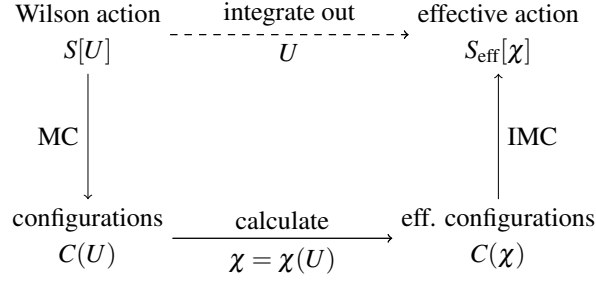
\begin{figure}[htb]
\centering
\scalebox{0.9}
{
\begin{tikzpicture}[every text node part/.style={align=center}]
\node (A) at (0,0) {Wilson action \\ $S[U]$};
\node (B) at (6,0) {effective action \\ $S_{\text{eff}}[\chi]$};
\node (C) at (0,-3){ configurations \\ $C(U)$};
\node (D) at (6,-3){eff. configurations \\ $C(\chi)$};

\draw[->,dashed] (A) -- (B) node [pos=0.5,above] {integrate out};
\draw[->,dashed] (A) -- (B) node [pos=0.5,below] {$U$};

\draw[->] (A) -- (C) node [pos=0.5,left] {MC};

\draw[->] (C) -- (D) node [pos=0.5,above] {calculate};
\draw[->] (C) -- (D) node [pos=0.5,below] {$\chi=\chi(U)$};

\draw[->] (D) -- (B) node [pos=0.5,right] {IMC};
\end{tikzpicture}
}
\caption{Inverse Monte-Carlo-Method}
\label{imc}
\end{figure}

Since we cannot derive the actions of the effective Polyakov-loop
models (\ref{linansatz}) and (\ref{logansatz}) in closed form by
analytically integrating out the spatial links, we determine their coupling
constants via Inverse Monte-Carlo (IMC).  
In the IMC method we first generate configurations of the full theory via
Monte-Carlo, calculate the corresponding configurations in
terms of the degress of freedom of the effective theory,
and then use the latter to determine the couplings of the effective theory
in the IMC step (see Fig. \ref{imc}). 
In principle the IMC step is done by taking an ansatz for an effective
action $S_{\text{eff}}(\lambda)$ with yet to determine coupling
constants $\lambda$. As in the derivation of Dyson-Schwinger equations
(DSEs) we use that expectation values of total derivatives
with respect to the fields in the effective action must vanish and require  
that this remains true when replacing the effective theory with the
full theory for the calculation of these expectation values via
Monte-Carlo, i.e. 
   \begin{align}
0=\left<\frac{\delta S_{\text{eff}}}{\delta\varphi}(\lambda)\right>_{\text{eff}}\stackrel{!}{=}\left<\frac{\delta S_{\text{eff}}}{\delta\varphi}(\lambda)\right>_{\text{full}}. 
\label{dse}
   \end{align}
This requirement implicitly determines the coupling constants of the
effective theory. Since in lattice gauge theory  we are dealing with
link variables, which are elements of a gauge group $G$, in order to
derive a  DSE we need derivatives and integrations with
respect to group elements. From the left invariance of
the Haar measure, yielding for the left derivative $L_a$ of a function
$f$ on $G$,
  \begin{align}
 \int d\mu(g)(L_a f)(g)=0\,, \quad f\in L_2(G) \, , \label{wardid1}
\end{align}
one can derive a DSE from geometric Ward-identities \cite{wardid}.  
Setting the function $f$ to $\vec L \cdot(F\vec{L}\tilde F)$, with
class functions $F$ and $\tilde F$,  
which then itself is a class function likewise, and using the fact that
the fundamental characters $\chi_q$, with
$q\in\{1,\dots,r=\mathrm{rank}(G)\}, $ 
provide a basis for class functions, we can apply a character expansion
 \begin{align}
 &  L_a F(\mathbf \chi)=\sum_{q=1}^r \frac{\partial F(\mathbf
     \chi)}{\partial \chi_q(g)} \, L_a\chi_q(g) \, . 
 \end{align}
 Setting $\tilde F = \chi_p$, for some  $p\in\{1,...,r\}$, one derives
 the master equation 
 {\begin{align} \label{wardid3}
  &0=\int_G d\mu_{red}\left\{  \frac{1}{2}\sum_{q}K_{pq}\frac{\partial F(\mathbf \chi)}{\partial \chi_q(g)} - c_p \chi_p(g)F\right\}, 
  &K_{pq}:=\left[(c_p+c_q)\chi_p\chi_q -\sum_\rho C^\rho_{pq}c_\rho\chi_\rho \right],
\end{align}}%
where  $C^\rho_{pq}$ are Clebsch-Gordon coefficients, $c_\rho$
eigenvalues of corresponding Casimir operators,  
and the sum runs over all irreducible represenations $\rho$.
One equation of the form (\ref{wardid3}) holds independently for
every point on the $d-1$ dimensional lattice of our effective theory. 
Therefore, inserting unity in terms of $\Pi=\exp(-S_{\text{eff}})$
times its inverse, one can write the lattice average of these
equations in the form of expectation values. In the last step we
replace the measure for these expectation values with
that of the full theory,
\begin{align} \label{geomdse}
 V^{-1}\sum_{i\in L}\left<\frac{1}{2}\sum_q K_{pq,i}\frac{\partial
   \vec F_i}{\partial \chi_{q,i}}\Pi(\vec
 \lambda)^{-1}-c_p\,\chi_{p,i} \,\vec
 F_i\,\Pi(\vec\lambda)^{-1}\right>_{\text{full}}=\vec 0. 
\end{align}
Moreover, we have collected sets of as yet unspecified class functions per
lattice site $i$ in large vectors $\vec F_i$ because their number needs to
match that of the couplings in the ansatz for the
effective action, i.e.  $\dim(\vec F_i)=\dim (\vec\lambda)$, so that
the resulting system of DSEs (\ref{geomdse}) can be solved to
determine the couplings $\vec \lambda $ via IMC. 

For $SU(2)$ there is only one fundamental represenation, with $p=q=1$
and $c_1=3/2$, $c_3=4$, $C^3_{11}=1$ in Eq.~(\ref{wardid3}). By
setting the functions  $\vec F_i\equiv \vec f_i \Pi$  with
$\Pi = \Pi_{\text{lin}}$ for the linear model (\ref{linansatz}) and  $\Pi=
\Pi_{\text{log}} $ for the logarithmic model (\ref{logansatz}), respectively,
most of the factors $\Pi^{-1}$ in (\ref{geomdse}) cancel. A convenient choice
for the functions $\vec f_i$ in either case is then obtained from  
 \begin{align}
  f_{l,d,i}&=\frac{1}{\lambda_{l,d}}\frac{\partial \ln
    (\Pi_{l,d,i})}{\partial
    \chi_{1,i}},\hspace{1cm}\mbox{and} \hspace{1cm} 
f_{l,d,i}=\frac{1}{g_{l,d}}\frac{\partial \Pi_{l,d,i}}{\partial \chi_{1,i}},
 \end{align}
 with
 $(\vec f_i)_{d+ (l-1)\cdot r_{max}^2 } = f_{l,d,i}$ for the quadratic distance
 $d\in\{1,\dots,r_{max}^2\}$ and the represenation $l\in\{1,\dots,p_{max}\}$. It avoids
 any coupling constant from only occuring in terms that contain odd
 powers of some Polyakov loop  $\chi_{l,i}$ which would otherwise lead
 to an independence of Eqs.~(\ref{geomdse}) on that coupling after  
 group integration.

\section{Results}
For the linear model all dependences on the coupling constants are
linear and Eq.~(\ref{geomdse}) reduces to a matrix equation of
expectation values which can be solved by simple matrix inversion.  
In the logarithmic case we solve the then non-linear
Eq.~(\ref{geomdse}) by applying a secant method. 
Doing so we will investigate the linear and logarithmic model with
differnt values of $r_{max}^2$ up to 81, and $p_{max}$ up to Dynkin
label 3, i.e.~including the fundamental, the adjoint and the $4$-dimensional
representation, so that the maximum number of independent couplings is given by
$|\{\lambda_{l,r^2}\,: \hspace{.3cm}
r^2=1,\dots,r_{max}^2, \hspace{.1cm}
l=1,\dots,p_{max}\}|=r_{max}^2\cdot p_{max} = 243 $.

\subsection{Linear vs. Logarithmic Polyakov-loop Models}

First we compare the linear and the logarithmic models with only
nearest neighbour interactions.  
\begin{figure}[htb]
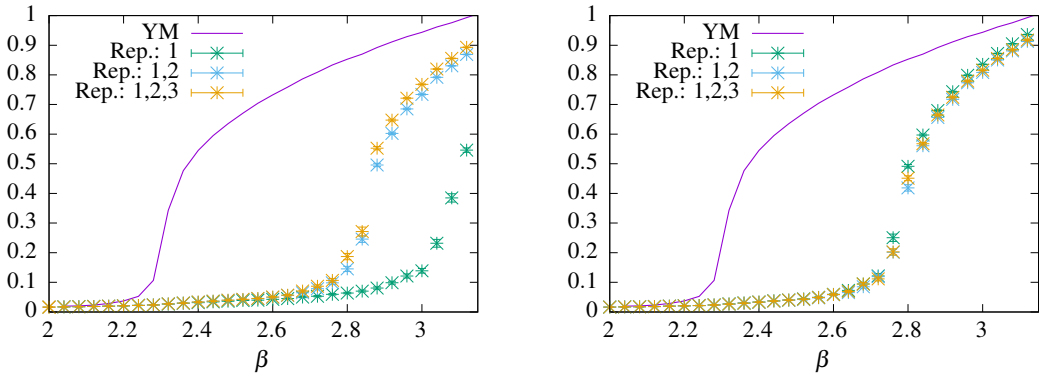

\begin{center}
 \scalebox{0.8}{\input{largerange_lin.tex}}\hskip 10mm
 \scalebox{0.8}{\input{largerange_log.tex}}
\end{center}
\vspace{-.8cm}
\caption{Polyakov-loop expecation values in local ($r_\text{max}=1$)
  linear (left) versus logarithmic (right) models compared to the $SU(2)$ gauge theory
  (YM) on a $16^3\times 4$ lattice.}
\label{locallinlog}
 \end{figure}
 As shown in Fig.~\ref{locallinlog}, the  linear model significantly
 improves the expectaion value of the Polyakov loop around criticality
 if we add up to 3 representations, while the 
 logarithmic resummation seems to work equally well with only the fundamental
 representation already.  
 Adding higher representations does not seem to improve the result much. 
 Below the phase transition, where the strong coupling expansion is
 valid, the logarithmic resummation is expected to improve the
 results. For $\beta \sim \beta_\text{c}$ there is still a large discrepancy
 between the best local model and the full theory.  
 We therefore gradually increase the non-locality up to
 $r_\text{max}=9$ to improve the logarithmic model.
 \begin{figure}[htb]
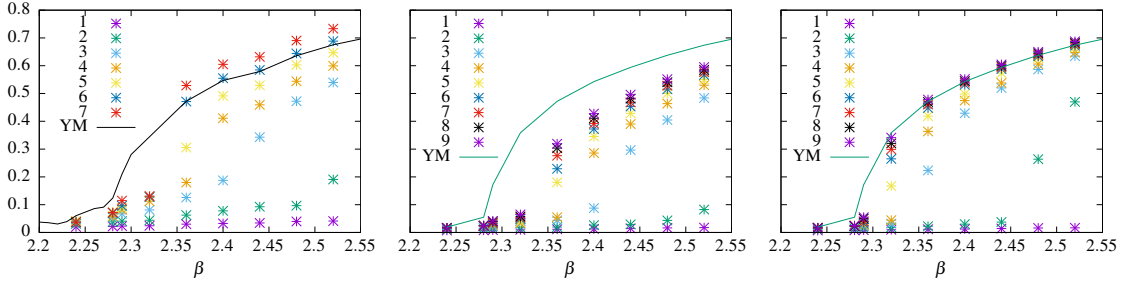

    \scalebox{0.6}{\input{phasendiagramm_log_16_1.tex}}
    \scalebox{0.6}{\input{phasendiagramm_log_32_1.tex}}
    \scalebox{0.6}{\input{phasendiagramm_log_32_2.tex}}
  \caption{Logarithmic model with non-local couplings on a
    $16^3\times 4$ lattice and fundamental representation only
    (left), and on a $32^3\times 4$ lattice with one (middle) and two
    representations (right).} 
     \label{nonlocallog}
   \end{figure}
 The expectation value of the Polyakov loop thereby indeed moves closer to that of
 the full theory near $\beta_\text{c} \approx 2.29 $ at first,  but
 evetually overshoots the gauge-theory result at larger $\beta $, as
 seen in the left panel of Fig.~\ref{nonlocallog}. 
 Larger lattices appear to fix this problem at first, but adding
 higher representations makes the result worse again (the more
 non-local couplings, the more the higher represenations start to matter
 again) as can be seen in Fig.~\ref{nonlocallog} (middle and right).  
 The model does not seem to converge in any clear way towards the full theory. 
 This might not be too suprising since the logarithmic resummation relies on
 the strong-coupling expansion which breaks down for larger $\beta$.
 Since the linear ansatz (\ref{linansatz}) is more general, one might
expect it to yield better results at large $\beta$. 
Again we now investigate the non-local linear Polyakov model by
gradually increasing $r_\text{max}$ up to 9.
 \begin{figure}[htb]
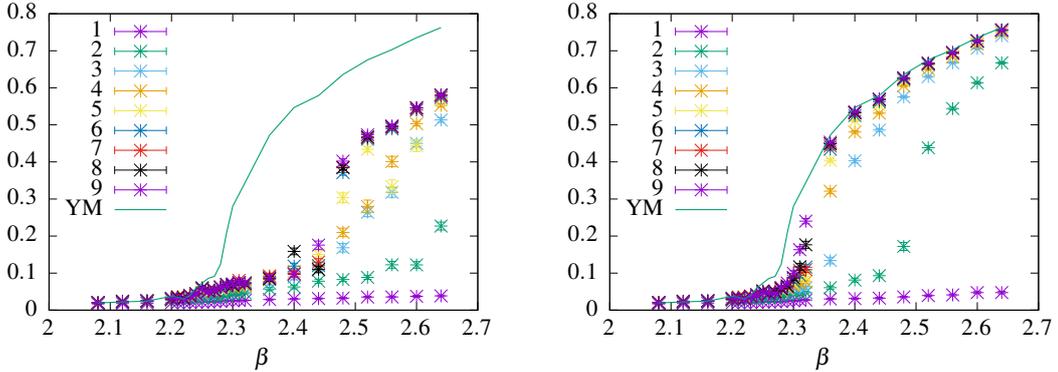

  \begin{center}
 \scalebox{0.8}{\input{phasendiagramm_lin_16_1.tex}}\hskip 10mm
 \scalebox{0.8}{\input{phasendiagramm_lin_16_2.tex}}
 \end{center}
 \vspace{-.8cm}
  \caption{Non-local linear model with one (left) and two representations
   (right) on a $16^3\times 4$ lattice.} 
 \label{nonlocallin}
 \end{figure}
As seen in Fig.~\ref{nonlocallin} for the $16^3\times 4$ lattice
the linear model is impoved by increasing $r_\text{max}$ without the
overshooting at large $\beta$. Moreover, adding the adjoint
representation to the ansatz improves the result significantly and we
match the full theory quite well in a wide range of $\beta$ around
$\beta_\text{c} $. We have checked that adding higher representations
does not change these results anymore. The linear model is more stable
than the logarithmic one when improving the truncation and provides
the better approximation to the full theory at larger
$\beta$. Nevertheless the model still approaches the full theory very 
slowly near $\beta_\text{c} $ indicating that one
might need to increase $r_\text{max}$ much further in this region (at
the expense of larger spatial volumes) which might reflect the
diverging correlation length of the theory at criticality. We have verified these results on the $32^3\times 4$ lattice as well.
 
\subsection{Long-Distance Behaviour of Non-Local Couplings}

\begin{figure}[htb]
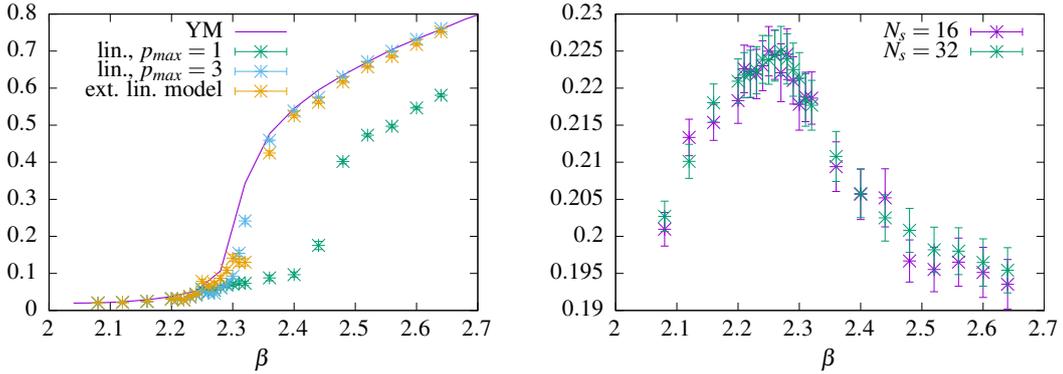

\begin{center}
  \scalebox{0.8}{ \input{polyakov_compare.tex} }\hskip 10mm
 \scalebox{0.8}{\input{lambdacorrelator.tex}}
\end{center}
\vspace{-.8cm}
\caption{Polyakov loop in the non-local linear model with $p_{max}=1$ and $3$
  compared to that in the extended linear model with potential term (left); 
  and characteristic length  $\xi(\beta)$ for the exponential fall-off of the
  couplings $\lambda_{1,r^2}$ with $r$ as a function of $\beta$ for the
  $16^3\times 4$ and $32^3\times 4 $   lattices (right).} 
\label{greensite}
 \end{figure}

In this section we look at the long-distance behaviour of the
couplings and compare it to  
a semi-analytical model proposed by Greensite and Langfeld \cite{greensite1}, 
given by the effective action
\begin{align}
 S_{GL}=-\frac{1}{8}\,c_1\,\sum_x
 \chi_x^2+\frac{1}{2}\,c_2\,\sum_{x,y}{}\chi_x \, Q(x-y)\,\chi_y \, ,
 \label{glansatz}
\end{align}
where $Q(x-y)$ is the square root of the negative Laplacian, i.e.
\begin{align}
 Q(x-y)=\begin{cases}
         (\sqrt{-\Delta_L})_{xy} & |x-y|\leq r_{max}\equiv 3\ \\
         0  &|x-y|> r_{max}\equiv 3
        \end{cases}.
\end{align}
On a $16^3\times 4$ lattice with $\beta=2.22$ the constants were
given as $c_1\approx 4.417(4)$ and $c_2\approx 0.498(1)$. 
 Comparing Eqs.~(\ref{glansatz}) and (\ref{linansatz}), we see that
 the effective theories are very similar if we use
 $p_{max}=1$, add a quadratic Polyakov loop potential  
with coupling constant $\lambda_{1,0}$ (which yields an extended linear model), and apply the mapping
\begin{equation}
 \lambda_{1,0}=-\frac{1}{8}\,c_1+\frac{1}{2} \,c_2\,Q(0) \quad
 \text{and} \quad \lambda_{1,r^2}=c_2 \, Q(r) \quad \text{with}\quad
 Q(r)\approx\frac{1}{|\{x : |x|=r\}|}\sum_{|x|=r} Q(x). 
\end{equation}
A fit to our couplings obtained from IMC then leads to
$c_1=3.6(7)$ and $c_2=0.42(7)$, consistent  with the values given by
Greensite and Langfeld.
In Fig.~\ref{greensite} (left) we compare the expectation value of
the Polyakov loop in the extended linear model
to our best results ($r_\text{max} = 9$) with one and three
representations in the linear model (\ref{linansatz}). We observe
that the quadratic (adjoint) potential has almost the same effect as
adding non-local interactions in the adjoint and higher representations.
 Close to the phase transition we have to take into account larger
 distances in the effective action. 
 In order to quantify this behaviour we define a characteristic length
 $\xi(\beta)$, via 
 \begin{equation}
  \lambda(r) \propto a(\beta)\cdot\exp(-\frac{r}{\xi(\beta)}) \, ,
 \end{equation}
 and check for critical scaling.
 Fitting this exponential fall-off to the first couple of
 IMC couplings at small $r$ for various $\beta$-values yields the
 results shown in Fig.~\ref{greensite} (right).
 While this characteristic fall-off does peak at around
 $\beta_\text{c}\approx 2.29$, the height of this peak does not grow
 with the volume as one would  expect for a correlation length near
 criticality. On the other hand, this might indicate that the theory
 becomes local again in the thermodynamic limit. To confirm this we have to
 investigate larger $N_t$ and extrapolate towards the contiuum limit,
 however.  
   
\section{Conclusion and Outlook}

In this contribution we have presented the IMC method for different
non-local pure gauge SU(2) Polyakov-loop models and saw that the
methods works quite well. We found that the logarithmic resummation is
only applicable for $\beta < \beta_c$, whereas for $\beta>\beta_c$ we
can approximate the full theory, in a way that shows clear signs of
convergence, only using the non-local linear Polyakov-loop models. We
also compared our non-local extended linear ansatz to the model
proposed by Greensite and Langfeld and found quite good
agreement. As expected, around the phase transition 
both classes of models (linear and logarithmic) approximate the full theory
only relatively poorly, however,  and we would need to include even
more non-local terms for a better description. On the other hand,
we also found indications that the linear model might become local in
the contiuum limit again. We have also computed more complicated
observables, like local Polyakov-loop distributions and 
correlation functions which we will present in an
upcoming full article. We will also investigate the behavior of our
models in the contiuum limit in the future, investigate higher rank
gauge groups such as $G_2$ and $SU(3)$, and add fermionic interactions
with finite chemical potential as relevant for the phase diagram of QCD.

\end{document}